\documentclass[conference]{IEEEtran}
\IEEEoverridecommandlockouts
\usepackage{cite}
\usepackage{amsmath,amssymb,amsfonts}
\usepackage{algorithmic}
\usepackage{graphicx}
\usepackage{textcomp}
\usepackage{xcolor}
\def\BibTeX{{\rm B\kern-.05em{\sc i\kern-.025em b}\kern-.08em
    T\kern-.1667em\lower.7ex\hbox{E}\kern-.125emX}}
\begin{document}

\title{The Approach to Managing Provenance Metadata and Data Access Rights in Distributed Storage Using the Hyperledger Blockchain Platform
\thanks{This work was funded by the Russian Science Foundation 
(grant No 18-11-00075).}
}

\author{\IEEEauthorblockN{Andrey~Demichev}
\IEEEauthorblockA{\textit{Skobeltsyn Institute of Nuclear Physics} \\
\textit{Lomonosov Moscow State University}\\
Moscow, Russia \\
demichev@theory.sinp.msu.ru}
\and
\IEEEauthorblockN{Alexander~Kryukov}
\IEEEauthorblockA{\textit{Skobeltsyn Institute of Nuclear Physics} \\
\textit{Lomonosov Moscow State University}\\
Moscow, Russia \\
kryukov@theory.sinp.msu.ru}
\and
\IEEEauthorblockN{Nikolai~Prikhod’ko}
\IEEEauthorblockA{\textit{Yaroslav-the-Wise} \\
\textit{Novgorod State University}\\
Velikiy Novgorod, Russia \\
niko2004x@mail.ru}
}

\maketitle

\begin{abstract}
The paper suggests a new approach based on blockchain technologies and smart contracts to creation of a distributed system for managing provenance metadata, as well as access rights to data in distributed storages, which is fault-tolerant, safe and secure from the point of view of preservation of metadata records from accidental or intentional distortions. The implementation of the proposed approach is based on the permissioned blockchains and on the Hyperledger Fabric blockchain platform in conjunction with Hyperledger Composer.
\end{abstract}


\section{Introduction}
Currently, the flow of data from a variety of sources, such as various sensors, WWW, mobile communication devices, large scientific experiments, etc., is growing at an enormous rate. In particular, modern scientific experimental facilities from the category of megascience, for example, LHC (Large Hadron Collider; http://www.cern.ch), LSST (Large Synoptic Survey Telescope, http: //www.lsst .org), are faced with the need to work with unprecedented amounts of data, to the brink of the opportunities offered by the computer industry and information technology. Only one of the detectors of the Large Hadron Collider, namely ATLAS, produces about a petabyte of raw data per second. Huge data sets are also generated in such experimental studies as the recording of meteorological data and astronomical observations.

As other examples of sources of big data, one can indicate continuously arriving data from various measuring devices, events from radio frequency identifiers, messages from social networks, meteorological data,  Earth's remote sensing, data streams about the location of subscribers of cellular networks, data from devices of audio and video registration. The development and the beginning of wide use of these sources initiates the active penetration of big data technologies into research and development, as well as in commercial and public administration.

Awareness of these changes in science and business has led to the understanding that it is necessary to develop new architectures and the principles of the operation of information systems in order to cope with this huge flow of data. Among other things, this applies to data storage systems. In existing centralized solutions, the main functions are performed by data centers that collect and store (possibly with subsequent processing) data from peripheral (user) nodes of the network. Therefore, in this case, the need to store big data in any scientific or production area leads to the necessity to build large and very expensive specialized data centers. At the initial stage of implementing a project, it is very problematic both to find sufficient funding for the establishment of such a center, and to estimate in advance the necessary storage capacity for a sufficiently long period of time.

The approach based on the peer-to-peer (P2P) paradigm of storage networks (see the review~\cite{1} and references therein) is totally opposite to the completely centralized approach. In this case, data storage services are evenly distributed among all network participants, which provides a natural load balancing, the absence of bottlenecks and points of failure. Special mechanisms of coding, fragmentation and distribution of information over nodes can provide privacy and reliability of the system even in case of failure of some storage nodes. However, a significant problem with this approach is to ensure a stable pool of peers, that is storage resource providers, especially at the initial stage of development of such a network. In other words, before such a P2P-based storage can work stably, it will require significant technical, organizational and time costs from its organizers in the absence of a result guarantee, that is, a workable network with sufficient storage capacity.

In many cases, the solution that is intermediate between a fully centralized and a fully decentralized (P2P) solutions may be optimal. For the implementation of such a solution, the organizations participating in a large project integrate their local storage resources into a unified distributed pool and, if necessary, rent in addition cloud storage resources, perhaps from multiple providers.  From an economic and technical point of view, such a solution may be particularly advantageous if there is a need to store large amounts of data for a limited duration of a project and in a situation where the project brings together many organizationally unrelated participants. In general case, such a distributed storage pool creates a dynamically changing environment (new storage can be connected to the pool or previously included storage can be disconnected as needed).

Each storage that enters this pool can have its own data management system (DMS). The problem is to combine all these data storages into a single system in a dynamically changing environment, and also ensure the implementation of reciprocal access policies to the data of the parties involved (for example, the term of paid lease of some (cloud) resource may expire, some organization can leave the project and stop access to their data, etc.). This implies the development of decentralized management methods both for data access rights in such a dynamically changing environment and for ensuring a reliable, immutable record of the history of committed transactions, that is, provenance metadata (PMD), for investigation and resolving possible conflicts between project participants, as well as owners of the storages. In other words, it is necessary to provide tools to support the implementation of business processes for storing and exchanging scientific data in a distributed environment and with administratively unrelated or loosely connected organizations participating in joint projects or simply exchanging data on certain conditions (for more details, see Section~\ref{PSBP}).

In this paper, we propose an approach to solving this problem based on the use of blockchain technology and smart contracts within the Hyperledger platform (https://www.hyperledger.org) \cite{2}. The basic principles of operation, architecture and algorithms of the ProvHL system (Provenance HyperLedger) for managing provenance metadata and data access rights in distributed storages are developed.

It should be noted that although a number of projects have been implemented in recent years to create systems for supporting and managing metadata, including the provenance of data, but the vast majority of the implemented solutions are centralized~\cite{3,4}, which is poorly consistent with the use of a distributed dynamically changing environment, and the possibility of using metadata by organizationally unrelated or weakly related research communities. On the other hand, in recent years, distributed registries based on blockchain technology have become very popular in various applied areas due to a number of important advantages~\cite{5,6}. Most recently, on the basis of the blockchains, developments have also been developed for the PMD management systems~\cite{7,8}. However, they are designed to work with one storage, do not solve the problem of providing a business process for data exchange between administratively different organizations and managing access to data.

\section{Managing the processes of storage and data exchange using the Hyperledger blockchain platform \label{MPSD}}

\subsection{Problem statement: business processes of data storage and exchange in a distributed environment \label{PSBP}}

The basic scenario of using the proposed system assumes that a virtual organization (VO) is formed for the joint implementation of a certain project. VO includes several real organizations, in turn including data providers, users and data handlers affiliated with them. It is assumed that the implementation of such a project requires the use of a distributed data storage. This distributed storage can be formed by renting multiple cloud storage, as well as integrating the own storage resources of the organizations that form the VO. Thus, the hardware and software basis of the business environment in this case is formed by a set of storages (possibly of different types, e.g., cloud storages, file servers, tape storages, etc.), each of which can be managed by its own data management system (DMS). For certainty, it is further assumed that the data is stored as files, i.e. the file is a unit of data. Generally speaking, several VOs can coexist; the storages with which they interact can form partially overlapping sets.

In such an environment, an immutable and distributed (as the environment itself) registry and a consensus on the order of data operations are needed to resolve possible conflicts between the VO/project participants related to the use of the data. In other words, to resolve possible conflicts, one needs an undistorted history of data use by VO members. Conflicts may be caused by priority issues upon obtaining results of data processing, use  of results (including funding issues), etc.

The state of the data (file) is determined by its provenance metadata (PMD), which consist of its global file identifier (ID) and its attributes, including:
\begin{flushleft}
\begin{enumerate}
\item  local file name in a storage: fileName;
\item storage identifier: storageID;
\item creator identifier: creatorID;
\item owner identifier: ownerID
\item type: type=primary/secondary/replica
\item source: source=\{expFacility,expFacilityID\}/ \{\{file,fileID\},\{tool,toolName\}\}
\item date/time of creation: dateTime=\{yyyy.mm.dd;hh.mm.ss\}
\item number of file downloads: downloads=integer 
\item users who downloded the file: dUsers=\{\{userID,txID\}, \{userID,txID\},\dots,\{userID,txID\}\}
\item URI of file description: metadataURI 
\end{enumerate}
\end{flushleft}
In step 6, the {tool, toolName} pair refers to the tools with which the file is obtained from the source.
The set of values for all attributes of a file determines its current state. The state of the entire distributed storage system is determined by the set of files stored in it with their states at the moment.

Basic transactions can be of the following types:
\begin{itemize}
\item new file upload;
\item file download;
\item file copy within a storage;
\item file deletion;
\item file copy to another storage;
\item file transfer to another storage.
\end{itemize}
Each active transaction corresponds to an update of some state keys, for example, after the transaction ``file download'' the values of the keys change: ``number of file downloads'' and ``users who downloaded the file''.

In addition to the task of recording the immutable history of working with data in a distributed storage environment, the task of providing distributed management of access rights to data is set. For example, the owner of a data file (the user who created the data, the organization to which it belongs --- through its authorized representative) must be able to manage its access rights for other users. Another example is when a cloud storage service grants access to data stored on it only to users from organizations that have paid for this storage service.

\subsection{Hyperledger blockchain platform \label{HBCP}}
A natural solution for the establishment of a distributed immutable registry for the PMD records is the use of the blockchain technology. To implement this solution, it is convenient to use existing blockchain platforms. Analysis of existing platforms shows that the tasks set in the previous section most naturally can be solved on the basis of the Hyperledger Fabric blockchain platform (HLF;  www.hyperledger.org)~\cite{2} together with Hyperledger Composer (HLC; hyperledger.github.io/composer). The latter is a set of tools for simplified use of the blockchain. The volume of this paper does not allow a detailed analysis and comparison of different platforms. As a partial filling of this gap, we refer to the existing comparative reviews~\cite{9,10,11}. Hyperledger Fabric together with Hyperledger Composer ((HLF\&C)-platform) has the following main advantages:
\begin{itemize}
\item it works within the concept of permissioned blockchains, in which transaction processing is carried out by a certain list of trusted network (distributed system) members; the resulted environment is more controlled and predictable than in the case of public permissionless blockchains, while the creation of blocks does not require resource-intensive calculations related to the proof-of-work algorithms~\cite{6};
\item it provides the operation of smart contracts (called chaincode), which allows us to organize the business process of sharing storage resources by project participants located in different administrative domains;
\item it has advanced means of managing access rights to certain actions, and access rights can be managed by network members within their competence (for example, the file owner can manage access rights to operations on them for other participants);  
\item it provides a record of transactions and advanced query tools concerning both the current state of the system and its evolution (history of transactions);
\item thanks to its modular structure, it allows using different algorithms to reach consensus between business process participants;
\item it allows for simultaneous independent operation of several virtual organizations, each within the framework of its project (channels in the terminology of HLF); 
\item it has a developed built-in security system based on PKI (public key infrastructure).
\end{itemize}
The umbrella-type project Hyperledger (https://www.hyperledger.org) was founded by the Linux Foundation in 2015 for the development of blockchain technology. Over time IBM, Cisco, Fujitsu, Hitachi, Intel, J. P. Morgan, SWIFT, Wells Fargo became participants of the project, in total more than a hundred large companies. One of the features of Hyperledger is the fundamental refusal to create their own cryptocurrency. The participants of the Hyperledger project develop it decidedly as information technology.

From a functional point of view, the nodes in the HLF network are divided into three types:
\begin{itemize}
\item clients make requests to execute transactions, participate in their processing, and broadcast transactions to ordering services;
\item peers carry out the transaction processing workflow, validate them and manage the blockchain registry;
\begin{itemize}
\item the blockchain registry is an append-only data structure, recording all transactions as a hash chain, as well as the succinct representation of the latest ledger state;
\item not all peers process each transaction; the subset of peers that process a given transaction is determined by the policy of the chaincode to which it belongs; peers from this subset are called endorsing peers or, simply, endorsers;
\end{itemize}
\item Ordering Service Nodes (OSN) or, simply, orderers  establish the general order of all transactions in the blockchain based on the distributed consensus algorithm; each transaction contains updates to the state of the system, the history of which contains the blockchain, as well as cryptographic signatures of endorsing peers; separation of processing nodes (peers) and transaction ordering makes the consensus in HLF as modular as possible and simplifies the replacement of consensus protocols.
\end{itemize}
To describe the business process within the framework of  (HLF\&C)-platform, a number of concepts are used, the main ones are assets, participants, transactions and events. Assets are tangible or intellectual resources, services or property, records of which are kept in registries. Assets can represent almost anything in a business network, such as a house for sale, a listing for sale, a land registration certificate for that house, while insurance documents for that house can be assets in one or more business networks. Assets must have a unique identifier, but they can also contain any properties defined for them.

In our case, the assets are data files; their properties (attributes) are provenance metadata, as defined in the previous Subsection~\ref{PSBP}. Participants are members of the business network. They can own assets and make transaction requests. Like assets, the participants must have an ID and can have any other properties if necessary. A participant can be associated with one or more identifiers. The transaction is the mechanism of interaction of participants with assets. The definition of events is also established in the process of  a business network construction, similarly to the assets or participants. According to these definitions, event messages can be sent by transaction processors to inform external software components of changes in the blockchain. Applications can subscribe to receive event information via the HLC API.

Unlike public blockchain networks, which allow non-authenticated users to participate in their work, members of the (HLF\&C)-network must be registered with Membership Service Provider (MSP), which, among other things, performs the functions of Certification Authority (CA).

It is worth noting that Hyperledger Fabric and Composer are in the stage of active development, so when using them one has to spend a lot of effort to fill the gaps in the extensive, cumbersome and yet not complete documentation of this platform.

\subsection{Basic principles of the ProvHL system operation \label{BPSO}}
This section outlines the basic principles of the ProvHL system for managing provenance metadata and access rights to data in distributed storages based on the HLF\&C blockchain platform.

In general, two approaches are possible:
\begin{itemize}
\item data management systems (DMS) manage data and use blockchain as a distributed log (data driven data management);
\item first, the metadata is written to the blockchain, and DMSs refer to the blockchain and performs the transactions recorded there (metadata driven data management).
\end{itemize}
In the first case, the functionality of the blockchain system is very limited,  it only provides resistant to malicious attempts to modify the history of data in distributed storage. (HLF\&C)-platform enables one to implement the second approach, which in addition to simply maintaining the registry allows us to solve the problem of distributed data access management, which is formulated in section II.A. Note that the term ``metadata driven'' is most often used in the context of ETL-technology; we use it solely to designate a way to manage data by pre-writing metadata to the HLF.

Only provenance metadata is recorded and stored in the blockchain. However, provenance metadata can refer to descriptive metadata (descriptive metadata is stored in a separate database).

\begin{figure}[htbp]
\centerline{\includegraphics[width=0.45\textwidth]{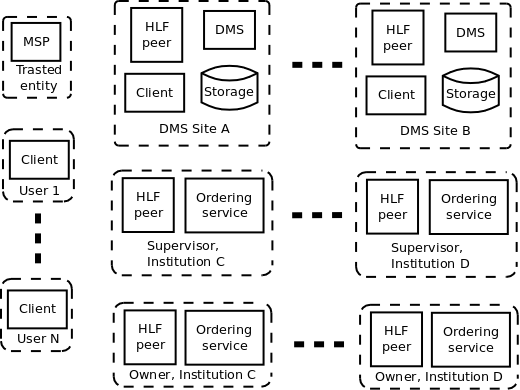}}
\caption{Distribution of the main HLF software modules by the administrative domains of the distributed storage environment.}
\label{fig1}
\end{figure}

The distribution of the main HLF modules by administrative domains of the distributed storage environment within the ProvHL system is shown in figure~\ref{fig1}.

The figure shows the distribution of HLF nodes (software components), namely, clients, peers, ordering services by business process actors, that is, participants in the terminology of (HLF\&C): users, DMSs (for distributed repositories A, B,...), supervisors (for example, the administrative management of the real organizations C,D,...) owners (for example, scientific project management from real organizations C,D,...). MSP is deployed separately and is a trusted service for all actors.

\begin{figure*}[!t]
\centerline{\includegraphics[width=0.85\textwidth]{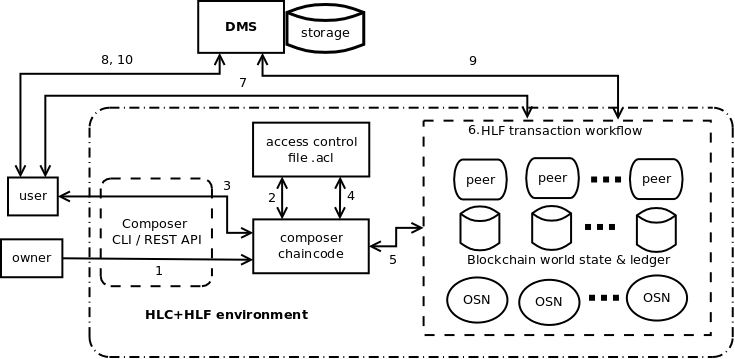}}
\caption{Simplified scheme for recording transactions with provenance metadata and managing data access rights based on HLF\&C.}
\label{fig2}
\end{figure*}

The interaction of the components of the (HLF\&C)-platform is not shown in figure~\ref{fig1}, it is shown in figure~\ref{fig2}. But it follows from the scheme in figure~\ref{fig1}, in particular, that DMSs, supervisors, owners participate in the approval of transactions. And supervisors and owners of different organizations are involved in the ordering (as already noted, this may be important, for example, in terms of priorities in obtaining any data or the results of their analysis).

Figure~\ref{fig2} shows a simplified scheme for recording transactions with provenance metadata and managing data access rights based on HLF\&C.

The arrows in the figure indicate the following actions:
\begin{itemize}
\item 1,2: the owner accesses the chaincode function, which, according to the acl-file (access control language), allows the owner of the data to grant access rights to these data to another user or group of users;
\item 3: a user who is granted access rights by the owner accesses the chaincode with a request to make a transaction (ClientRequest transaction) with data (for example, file download, upload or copy);
\item 4,5: chaincode verifies that such a transaction complies with the rules defined in the acl-file and, if it does, sends a request to the HLF environment to complete the transaction;
\item 6: HLF performs transaction processing (transaction workflow: simulation/ endorsements — ordering — validation — state updating);
\item 7: HLF sends a message (event) to the user about the successful transaction and its recording in the blockchain; the message also contains the transaction identification number;
\item 8: the user accesses the data management system (DMS) with a request to perform a data operation that contains the number of the corresponding transaction;
\item 9: DMS checks for a record of this transaction in the blockchain;
\item 10: if there is a record of the valid transaction, the DMS performs the required operation and, in turn, initiates a transaction record confirming that a data operation was performed (ServerResponse transaction).
\end{itemize}
As it can be seen, for each data operation, two transaction records are made in the blockchain: one corresponds to the client request (ClientRequest), and the second corresponds to the server response (ServerResponse).  In the simplified description of the algorithm, some details specific to certain types of transactions are omitted for brevity. In particular, when the ``new file upload'' transaction is performed, the transaction on creating the new asset, that is the data file, with the ``temporary'' label is first recorded in the blockchain. And only after the actual upload of the file in the storage, DMS initiates a transaction removing the label ``temporary'' and turns the uploaded file into a fully valid asset. Together with the above-mentioned splitting of transactions into the client and server parts, this makes the level of correspondence between the history recorded in the blockchain and the real history of the data in the distributed storage practically acceptable. As is known, ensuring full compliance of the real world history with the history recorded in a blockchain is outside the scope of the blockchain technology (the so-called the Oracle Problem, see e.g., https://medium.com/@DelphiSystems/the-oracle-problem-856ccbdbd14f).

\section{Conclusion \label{Conc}}
In this paper, we have formulated a general problem and functional requirements for the management system of provenance metadata and data access rights (Section~\ref{PSBP}), which can support the implementation of business processes for storing and exchanging data in a distributed environment, with administratively unrelated or loosely connected organizations, participating in joint projects, or simply exchanging data on certain conditions. Through the use of a new approach based on the integration of blockchain technology, smart contracts and metadata driven data management, the principles and algorithms of the system, entitled ProvHL (Provenance HyperLedger), are developed that are fault-tolerant, safe and secure management system of provenance metadata, as well as access rights to data in distributed storages. The problems of optimal choice of the blockchain type for such a system, as well as the choice of the blockchain platform are studied. Namely, it is proposed to use a permissioned type of blockchain and the Hyperledger blockchain platform, on the basis of which the ProvHL system is implemented.

At present, a polygon has been created on the basis of  SINP MSU, where a preliminary version of the ProvHL prototype is deployed to implement the developed principles and refine the algorithms of the system. In this preliminary version, a trivial consensus algorithm is used, in which the transaction recording order is determined by a single server (centralized orderer Solo in the terminology of HLF). However, in the future it is supposed to use full-fledged Byzantine fault tolerant consensus algorithms, in particular PBFT~\cite{12}.

The creation of ProvHL production level system will significantly improve the quality and reliability of the results obtained on the basis of processing and analysis of big data in a distributed computer environment.

\end{document}